\documentclass[twocolumn,aps,prl,superscriptaddress]{revtex4-2}

\usepackage{amsmath}
\usepackage{graphicx}
\usepackage{MnSymbol}
\usepackage{parskip}
\usepackage{makecell}

\begin{document}

\title{Role of transfer films and interfacial cracking in metallic sliding wear}

\author{R. Xu}
\email{r.xu@fz-juelich.de}
\affiliation{State Key Laboratory of Solid Lubrication, Lanzhou Institute of Chemical Physics, Chinese Academy of Sciences, 730000 Lanzhou, China}
\affiliation{Peter Gr\"unberg Institute (PGI-1), Forschungszentrum J\"ulich, 52425, J\"ulich, Germany}
\affiliation{MultiscaleConsulting, Wolfshovener str. 2, 52428 J\"ulich, Germany}

\author{B.N.J. Persson}
\affiliation{State Key Laboratory of Solid Lubrication, Lanzhou Institute of Chemical Physics, Chinese Academy of Sciences, 730000 Lanzhou, China}
\affiliation{Peter Gr\"unberg Institute (PGI-1), Forschungszentrum J\"ulich, 52425, J\"ulich, Germany}
\affiliation{MultiscaleConsulting, Wolfshovener str. 2, 52428 J\"ulich, Germany}

\begin{abstract}
The origin of wear particles in metallic sliding contacts remains debated. Classical views based on cold-welded junctions suggest that plastic yielding of the real contact area should lead to large wear coefficients, in apparent contradiction with the small values typically measured for metals. Here we argue that this discrepancy can be resolved if most junctions do not directly produce wear particles, but instead cause metal transfer and the formation of a weakly bound transfer film. Wear then occurs intermittently when fragments of this film detach due to crack propagation at the interface between the transfer film and the underlying bulk metal.

We perform unlubricated reciprocating sliding experiments on nominally smooth stainless steel, brass, and aluminum. For steel on steel, the wear mass loss shows an initial stage with negligible mass change up to a sliding distance of $\sim 2.4 \ {\rm m}$, followed by a linear regime. Transfer-film formation in dissimilar-metal contacts is evidenced by optical imaging, net mass gain of the steel slider, and energy-dispersive X-ray spectroscopy, and the collected debris is flake-like. These observations support a transfer-film-controlled wear mechanism associated with cold-welded junctions.
\end{abstract}

\maketitle

\setcounter{page}{1}
\pagenumbering{arabic}




\vskip 0.1cm

\begin{figure}[tbp]
\includegraphics[width=0.35\textwidth,angle=0]{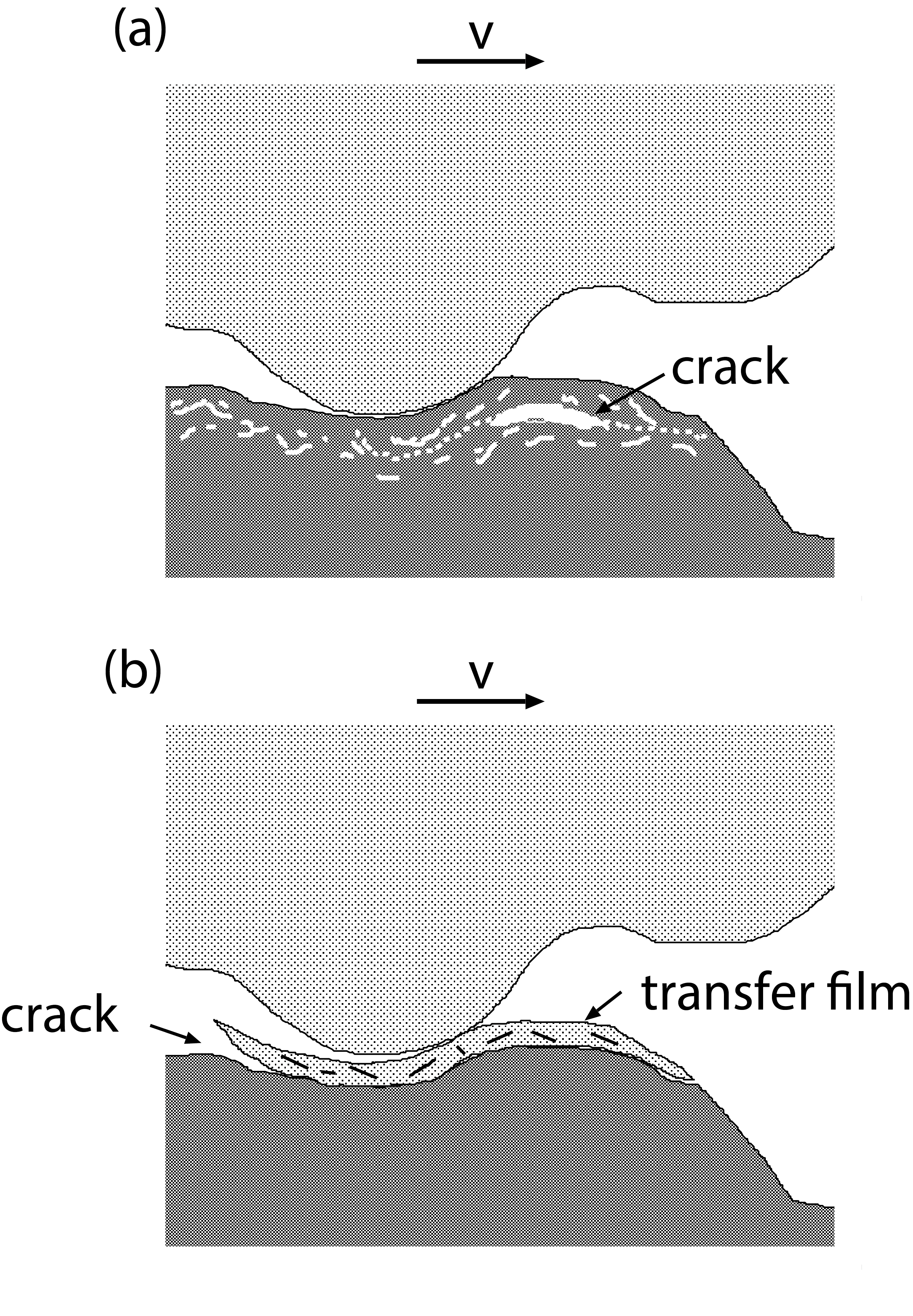}
\caption{
Schematic illustration of two wear mechanisms in metallic sliding contacts.
(a) Delamination-type wear where subsurface crack nucleation and propagation lead to the formation of platelet-like debris.
(b) Wear mechanism proposed in this study, where a weakly bound transfer film is formed by metal transfer at cold-welded junctions, and crack propagation at the interface between the transfer film and the underlying bulk metal causes fragments of the film to detach.
}
\label{PicRab1.eps}
\end{figure}

{\bf 1 Introduction}

In the classical book on friction and lubrication, Bowden and Tabor \cite{Bowden} presented a theory for the
friction between metals based on the formation and shearing of cold-welded junctions.
They assumed that when two metallic
blocks are squeezed together with a normal force $F_{\rm N}$, the contact occurs only in a small fraction of the
nominal contact area, where asperities make contact.
They further assumed that the local stress in the contact regions
is sufficiently large to plastically deform the metals.
For unlubricated metals, the
oxide and contamination films that are present on most metals in the normal atmosphere are broken or penetrated in the
contact regions, resulting in cold-welded junctions.

In 1986, Archard presented the following argument against the assumption that sliding contacts between metals yield plastically and form cold-welded junctions \cite{ArchComment}:
Assume that each cold-welded junction results in the removal of a particle with a linear size comparable
to the junction width $d$.
If there are $N$ junctions, then sliding a distance $L=d$ will generate $N$ wear particles
with volume $\Delta V \approx Nd^3$, so that
$${\Delta V \over L} \approx Nd^2.$$
If all contact has yielded plastically then the real contact area $A =Nd^2 = F_{\rm N}/\sigma_{\rm P}$, where $\sigma_{\rm P}$ is the penetration hardness,
and one obtains
$${\Delta V \over L} = K {F_{\rm N} \over \sigma_{\rm P}}\eqno(1)$$
with $K \approx 1$.
However, experiments for metals typically give
$K \approx 10^{-6}-10^{-3}$ (see Refs. \cite{A1,A2,Green,Rab1}), and Archard used this to argue that most contacts deform elastically rather than plastically,
which would reduce the formation of cold-welded junctions.

We argue that this conclusion is not compelling for nominally clean, unlubricated surfaces.
Archard's argument relies on the key assumption that each junction produces a wear particle.
In our view, most junctions formed in sliding contact do not directly generate wear debris.
Instead, they primarily lead to metal transfer between the surfaces and to the formation of thin metallic films that are bound more weakly to the substrate
than the bonding within the bulk metal.
From time to time, flake-like fragments detach from this transfer film and form wear particles.
Within this framework, there is no simple relation between the wear rate and the friction coefficient.
Rather, the wear rate depends on additional factors, such as the bonding strength of the transfer films and
the mechanisms by which fragments are removed from it.

This picture is related to the delamination theory of wear proposed by Suh \cite{nam} in 1973, where fatigue-driven nucleation and propagation of subsurface cracks
eventually produce platelet-like debris, as illustrated in Fig.~\ref{PicRab1.eps}(a).
Here we propose that, instead of subsurface cracks, material removal occurs via crack propagation at the interface between the transfer layer and the underlying bulk metal,
as illustrated in Fig.~\ref{PicRab1.eps}(b).

In this paper, we present reciprocating sliding experiments on several metallic systems to test this picture and to characterize the evolution of friction and wear
in relation to transfer-film formation and removal.

\vskip 0.3cm
{\bf 2 Experiments and results}

Specimen pairs made of various materials with nominally smooth surfaces were used.
The materials included standard AISI 304 stainless steel; a two-phase H59 brass alloy composed of
approximately 60 wt.\% Cu and 40 wt.\% Zn; and an AA 6061 aluminum alloy, a heat-treatable Al-Mg-Si alloy containing approximately 1.0 wt.\% Mg and 0.6 wt.\% Si.
The upper specimen is referred to as the slider, and the lower specimen as the substrate.
The sliders were prepared as square blocks measuring $10 \times 10 \ {\rm mm}^2$ with a thickness of $3 \ {\rm mm}$.
Chamfers of $1 \ {\rm mm}$ were made on all four edges to minimize edge effects during sliding, resulting in a nominal contact area of $8 \times 8 \ {\rm mm}^2$.
The substrates were flat plates with dimensions of $200 \times 100 \ {\rm mm}^2$.

Linear reciprocating sliding tests were conducted using a custom-built machine.
Each sliding cycle consisted of one forward and one backward motion of $120 \ {\rm mm}$ at room temperature ($\approx 30^\circ {\rm C}$).
The sliding speed was fixed at $1 \ {\rm mm/s}$ to minimize frictional heating.
The load was fixed at $180 \ {\rm N}$, resulting in a nominal contact pressure of $2.8 \ {\rm MPa}$.
Between test cycles, the slider and substrate were cleaned with anhydrous ethanol and nonwoven fabric to remove wear debris and contaminants.
Each test cycle was therefore considered to start from a clean surface.

During the experiments, the coefficient of friction and the wear rate of the slider were measured.
The coefficient of friction was recorded directly by the machine.
Mass loss was measured using a high-precision electronic balance (Huazhi PTX) with an accuracy of $0.1 \ {\rm mg}$.

\begin{figure}[tbp]
\includegraphics[width=0.45\textwidth,angle=0]{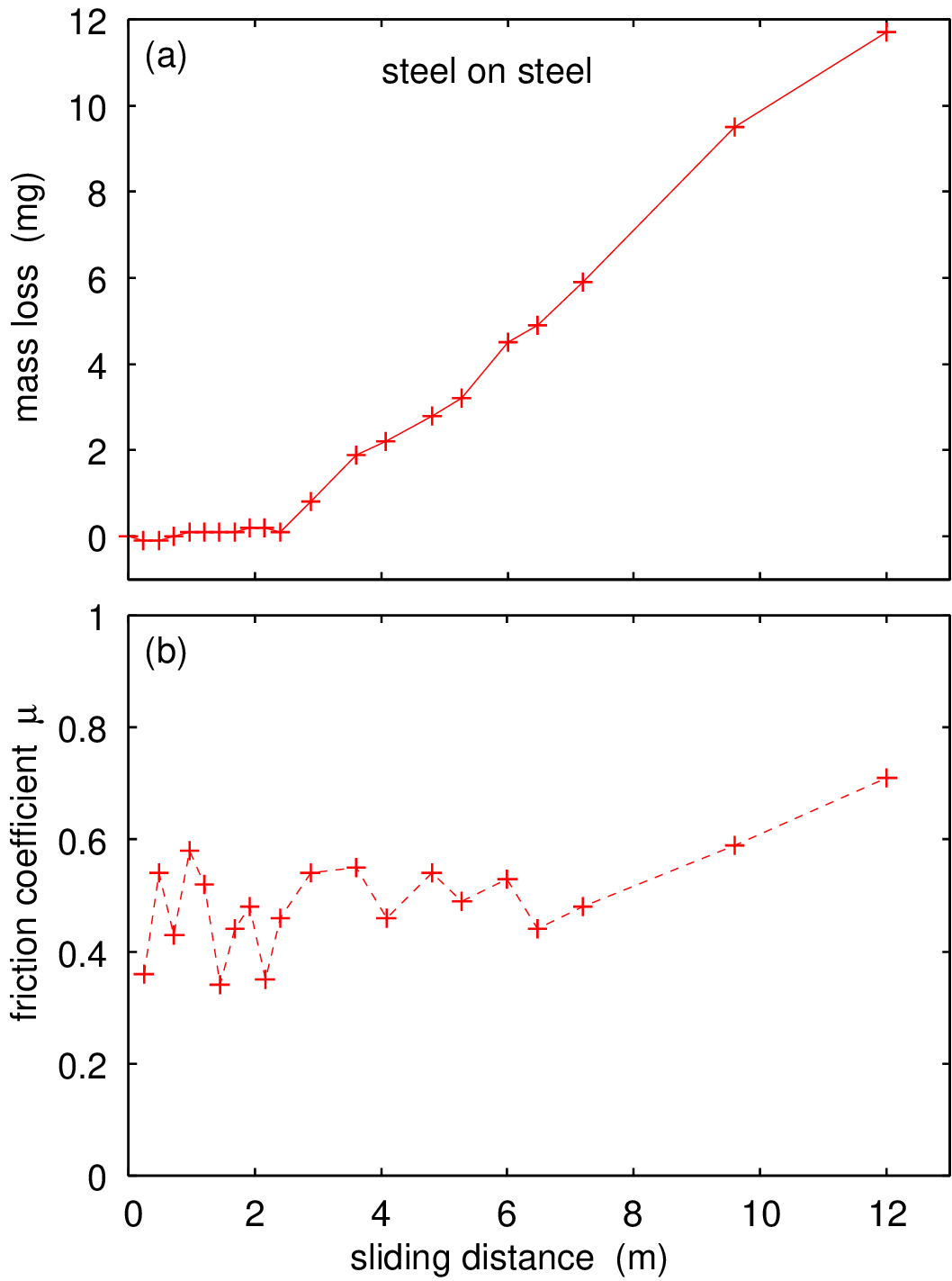}
\caption{
(a) Wear mass loss and (b) friction coefficient as a function of the sliding distance for a steel-on-steel contact.
}
\label{1distance.2massCHANGE2.eps}
\end{figure}

\begin{figure}[tbp]
\includegraphics[width=0.48\textwidth,angle=0]{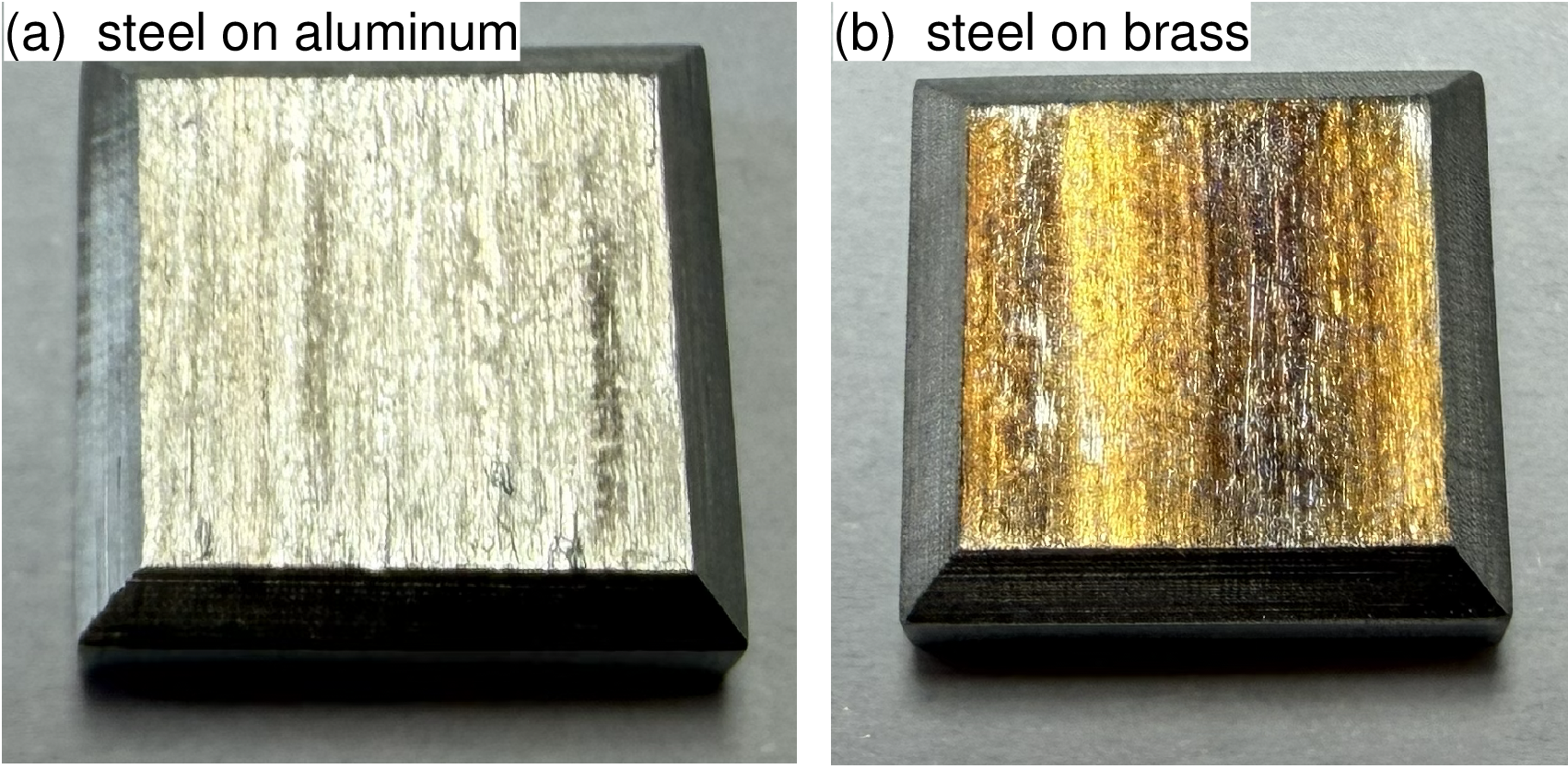}
\caption{
Optical images of a steel slider after sliding $64.8 \ {\rm m}$ on (a) an aluminum substrate and (b) a brass substrate.
For steel sliding on brass, the transfer film appears as yellow regions on the steel surface.
}
\label{SteelONaluminumandONbrass.eps}
\end{figure}

\begin{figure*}[tbp]
\includegraphics[width=0.9\textwidth,angle=0]{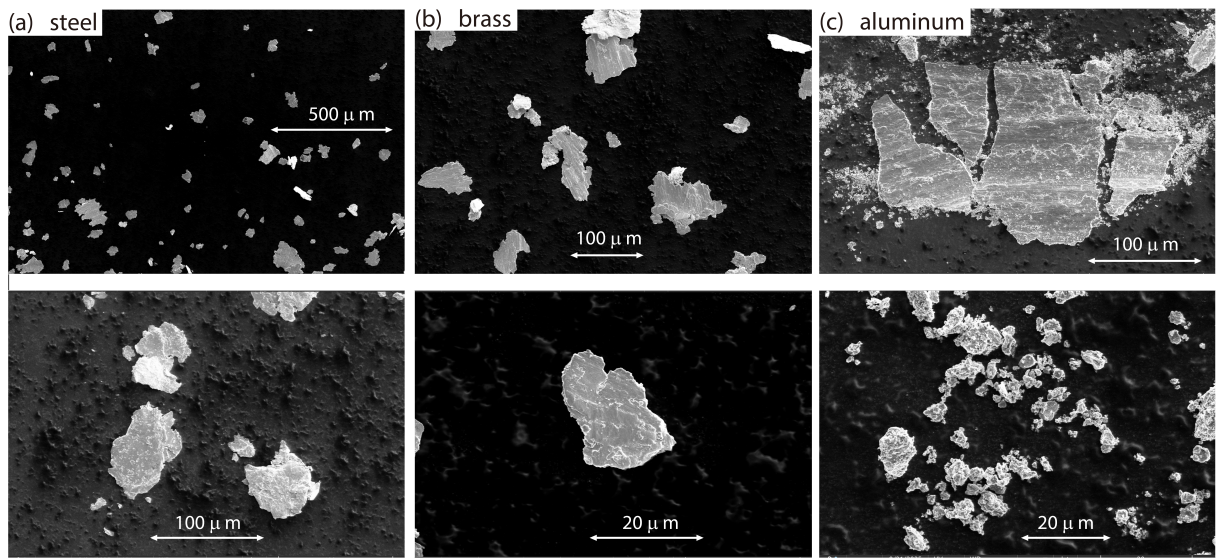}
\caption{
Scanning Electron Microscopy (SEM) images of wear particles collected from the substrate for identical-metal sliding contacts: (a) steel on steel, (b) brass on brass, and (c) aluminum on aluminum.
In all cases, the wear particles are flake-like, with lateral dimensions ranging from a few micrometers up to $\sim 100 \ {\rm \mu m}$.
}
\label{AllWearParticlesPicture.eps}
\end{figure*}

For a steel-on-steel contact, Fig.~\ref{1distance.2massCHANGE2.eps} shows (a) the wear mass loss and (b) the friction coefficient as a function of the sliding distance.
During the first $2.4 \ {\rm m}$ there is negligible wear.
After $2.4 \ {\rm m}$, the mass loss increases nearly linearly with the sliding distance.
In this linear regime, Eq.~(1) is obeyed with $K \approx 5 \times 10^{-3}$, where we have used
the penetration hardness for steel $\sigma_{\rm P} = 6 \ {\rm GPa}$.

Fig.~\ref{1distance.2massCHANGE2.eps}(b) shows that the friction coefficient remains nearly constant up to a sliding distance of $7.2 \ {\rm m}$.
For sliding distances longer than $12 \ {\rm m}$, the friction coefficient remains constant at $\sim 0.7$ (results not shown).
We also observed that after a sliding distance of $12 \ {\rm m}$, the surface roughness power spectrum no longer changes, which indicates that the run-in distance for the steel-on-steel system is $\sim 12 \ {\rm m}$.
The other metal systems studied show a similar run-in distance.
The surface roughness results are not shown in the present paper since here we concentrate on wear.

For sliding between identical metals, material transfer is difficult to observe directly, but for systems involving two different metals for the slider and the substrate,
metal transfer can be detected using energy-dispersive X-ray spectroscopy (EDS).
The EDS results show that metal transfer occurs between the steel slider and the aluminum and brass substrates.
For the latter case, this can also be seen in optical images, as shown in Fig.~\ref{SteelONaluminumandONbrass.eps}(b).
In addition, we observed an increase in the mass of the steel slider for dissimilar-metal contacts: after a sliding distance of $64.8 \ {\rm m}$, the steel slider mass increased by $0.7 \ {\rm mg}$ for steel sliding on brass and by $0.5 \ {\rm mg}$ for steel sliding on aluminum.
This net mass gain provides quantitative evidence for material transfer to the steel surface and suggests that the transfer process is strongly asymmetric for these pairs.

Wear particles collected from the substrate between identical-metal sliding contacts were studied using Scanning Electron Microscopy (SEM).
In all cases, we observe flake-like wear particles despite the obvious difference in particle sizes. This is shown in Fig.~\ref{AllWearParticlesPicture.eps}
for (a) steel on steel, (b) brass on brass, and (c) aluminum on aluminum.

\begin{figure}[tbp]
\includegraphics[width=0.35\textwidth,angle=0]{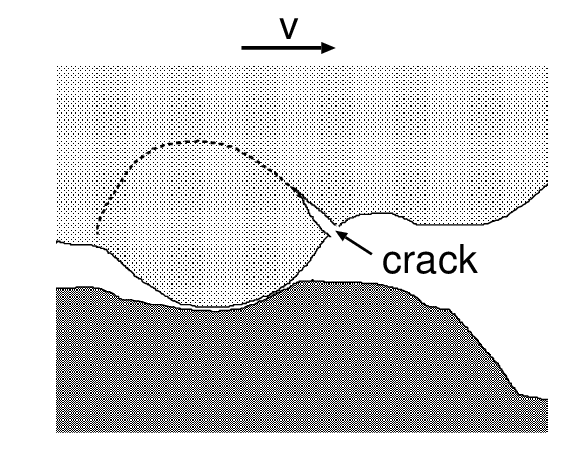}
\caption{
Energetic criterion for the formation of wear particles in asperity contacts.
To remove a particle of linear size $d$, sufficient elastic deformation energy must be stored in its vicinity to break the bonds required to form a free particle.
The bond-breaking energy scales as $U_{\rm c} \approx \gamma d^2$, where $\gamma$ is the cohesive energy per unit surface area, while the elastic energy scales as $U_{\rm el} \approx (\tau^2/E) d^3$, where $\tau$ is the shear stress in the asperity contact region.
Hence, particle removal is possible only if $d > \gamma E/\tau^2$, so only sufficiently large asperity contact regions give rise to wear particles.
}
\label{PicRab1Rabino.eps}
\end{figure}

\vskip 0.3cm
{\bf 3 Discussion}

The results reported above indicate that material transfer and the subsequent evolution of a transfer film play a central role in the present unlubricated experiments.
For steel on steel, the wear mass loss exhibits an initial stage with negligible net mass change followed by a regime of approximately linear mass loss, while the friction coefficient varies much less over the same sliding-distance range (Fig.~\ref{1distance.2massCHANGE2.eps}).
For dissimilar-metal pairs, transfer films are directly evidenced by optical imaging for steel sliding on brass (Fig.~\ref{SteelONaluminumandONbrass.eps}) and confirmed by EDS for steel sliding on aluminum and brass (results not shown).
The net mass gain of the steel slider for sliding on brass and aluminum further supports transfer-film formation.
In addition, SEM observations show that the collected debris is flake-like for all identical-metal contacts studied (Fig.~\ref{AllWearParticlesPicture.eps}).
Taken together, these observations support a picture in which wear is governed by the formation and intermittent removal of a weakly bound transfer film, rather than by the friction coefficient alone, and the wear process can be highly asymmetric.

The transfer of material from one surface to another during sliding has been the subject of numerous
studies \cite{Green}. By the late 1940s, radioactive isotopes were becoming more readily available to researchers, making it possible
to study material transfer processes in the wear of metals. These techniques were applied by Rabinowicz \cite{A4} in his studies of metal wear.
He presented a wear model for metals in which wear particles are removed when the elastic deformation energy stored in asperity contact regions becomes
larger than the energy needed to break the bonds between the particle and the surrounding material, as illustrated in Fig.~\ref{PicRab1Rabino.eps}.
In this model, the formation of a wear particle is due to crack propagation, and the energetic condition used by Rabinowicz is similar to the
fracture criterion formulated by Griffith \cite{griffith}.

One important implication of the Rabinowicz theory is that if the frictional shear stress
(or the kinetic friction coefficient) is small enough, there may not be sufficient elastic energy to
form wear particles. Hence, a boundary lubrication film can have
a large influence on the wear rate by reducing the frictional shear stress
to such a degree that insufficient elastic energy is available to form wear particles.
In this model, a strong dependence of the wear rate on the
sliding friction coefficient is expected, as was indeed observed
by Rabinowicz in a large set of experiments, where low wear rates and low friction occurred for lubricated contacts \cite{Rab1}.
In the present experiments, however, measurable wear for steel on steel appears after an initial stage, while the friction coefficient changes much less over the same sliding-distance range.
This suggests that, under the present conditions, wear is controlled primarily by the formation and detachment of a weakly bound transfer film, rather than by a direct one-to-one relation between wear rate and friction coefficient. 

The Rabinowicz-type energy criterion and related fracture-based removal mechanisms may be most relevant in the initial stage of wear for metals with very large roughness, and they may also apply to some brittle non-metals and to rubber-like materials \cite{RP1,RP2,RP3}.
For most nominally clean, unlubricated metallic surfaces, however, we suggest that the dominant wear mechanism involves fragments removed from weakly bound transfer films formed by metal transfer between the two sliding bodies due to cold-welded junctions.

\vskip 0.3cm
{\bf 5 Summary and conclusion}

We have investigated unlubricated reciprocating sliding for several metallic systems with nominally smooth surfaces, focusing on the relation between friction, wear, and material transfer. For steel on steel, the wear mass loss exhibits an initial stage with negligible mass change up to a sliding distance of $\sim 2.4 \ {\rm m}$, followed by a nearly linear increase. Over the same distance range, the friction coefficient changes much less, indicating that the wear rate is not determined by the friction coefficient alone under the present conditions.

For dissimilar-metal contacts, material transfer is directly evidenced by optical images for steel sliding on brass, and confirmed by EDS and by the net mass gain of the steel slider for steel sliding on aluminum and brass.
In addition, SEM observations show that the wear debris is flake-like for steel on steel, brass on brass, and aluminum on aluminum.

Based on these results, we propose that unlubricated metallic wear is a two-step process. First, metal transfer at cold-welded junctions forms a weakly bound transfer film. Second, when the transfer film becomes sufficiently developed, crack propagation at the interface between the transfer film and the underlying bulk metal leads to partial detachment of the film and the formation of flake-like wear particles. This mechanism is distinct from subsurface delamination, and it provides a physical interpretation of why small wear coefficients can coexist with junction formation in nominally clean metallic contacts.

\vskip 0.3cm
{\bf Data availability}

The data supporting the findings of this study are available from the corresponding author upon reasonable request.

\vskip 0.3cm
{\bf Acknowledgements}

We thank H.Ren (LICP, CAS, Lanzhou 730000, China) for the help with the SEM and EDS experiments.

\end{document}